*Andrey Y. Timashkov[1][2] , Ilya N. Abrosimov[3] , Vladimir M. Yaltonsky[3]*

# Illness perception and self-management in patients with type 2 diabetes

This paper presents the results of a study on the perception of illness and adaptation parameters in patients with type 2 diabetes. The study involved 173 patients diagnosed with "Type 2 Diabetes" (ICD-11 code 5 A 11). The average age of the patients was 55.21±13.47 and the average duration of the disease was 11.79±8.16. Two profiles of illness perception were identified: Profile 1 - "Perception of illness threat" and Profile 2 - "Perception of illness and treatment controllability". Three types of illness perception were also identified: Type 1 - "Formed illness threat and negative beliefs about illness and treatment control" (Group 1); Type 2 - "Unformed illness threat and neutral beliefs about illness and treatment control" (Group 2); Type 3 - "Formed illness threat and positive beliefs about illness and treatment control" (Group 3). Targets for further psychological interventions were formulated for each identified type.

JEL Classification: E39.

Keywords: type 2 diabetes, Illness perception**,** self-management, coping behavior, mindfulness, treatment adherence.

[1] Corresponding author. Email: aytimashkov@hse.ru
[2] Institute for Cognitive Neuroscience, HSE University; Moscow, 101000, Russian Federation.
[3] Faculty of Clinical Psychology, Moscow State University of Medicine and Dentistry; Moscow, 127473, Russian Federation.

# Introduction

According to the International Diabetes Federation, diabetes mellitus is one of the largest global health problems of the 21st century." In 2015, one in eleven adults was diagnosed with diabetes mellitus [4]. There are about 200 million diabetic patients in the world. In Russia for 2023, according to the national Federal register of patients with diabetes mellitus, 4.9 million patients (about 3.3% of the population) were registered, of which 5.58% were with type 1 diabetes, 92.33% with type 2 diabetes, 2.08% with other types of diabetes. [5]. According to statistics, every 10-15 years the number of people with diabetes doubles.

Such statistics indicate that diabetes mellitus is a global health problem with serious socio-economic consequences on a global scale.

Diabetes mellitus is a clinical syndrome of chronic hyperglycemia (increased blood sugar) and glycosuria (loss of glucose in urine) caused by absolute or relative insulin deficiency, leading to metabolic disorders, the development of micro– and macrovascular lesions, neuropathy and pathological changes in various organs and tissues [23].

At the moment, in relation to diabetes mellitus, a certain protocol of medical care has already been formed, aimed at maintaining optimal blood glucose levels in the patient, and there is also a treatment regimen for diseases associated with diabetes mellitus. Over time, treatment methods with a similar diagnosis are becoming more effective. However, to date, a similar protocol for psychological assistance to patients with diabetes has not yet been proposed. For a long time, this aspect was not given due attention, but as they immersed themselves in the problem, the importance of the psychological state of patients became obvious. Indeed, stress is one of the key factors in the occurrence of the disease, along with heredity and malnutrition, it is often isolated by patients themselves. Nevertheless, there are not enough studies on the psychological state of patients with diabetes mellitus and their attitude to the therapy provided in Russian and foreign publications.

To create a program of psychological care for patients with diabetes, it is necessary, first of all, to study the psychological state of these patients and their relationship with the disease. It is necessary to answer the questions: are there any common psychological features in these patients, and is it possible to classify the observed psychological phenomena? The answers to these questions will help us identify psychological care targets relevant to these patients, which in the future will become the basis for the formation of a psychological care program for people with diabetes.

This study is aimed at studying such psychological characteristics of diabetic patients as: parameters of coping behavior, attitude to the disease and emotional response to it, perception of social support, structure of awareness and commitment to pharmacotherapy.

In the study, we adhered to the following theoretical constructs: the concept of stress reduction based on mindfulness (J. Kabat-Zinn) [13]; concepts of the psychology of physicality and the internal picture of the disease (V.V. Nikolaeva, A.Sh. Tkhostov, G.A. Arina) [24]; cognitive-phenomenological theory of coping with stress (coping behavior) (Lazarus R.S., Folkman S.) [9,14], continued in the works of domestic specialists (Yaltonsky V.M., Sirota N.A., Sokolova E.T., Vasserman L.I., Isaeva E.R., Mikhailova N.F.) [2, 25, 26], including in the aspect of adherence to treatment (Yaltonsky V.M., Sirota N.A.) [22, 27].

**The aim of the study** is a comparative analysis of disease perception and adaptation parameters among patients with type 2 diabetes.

**Patients and methods.** The study involved 173 patients diagnosed with "Type 2 Diabetes" (ICD-11 code 5A11). Data from 4 patients did not pass the verification, thus the analysis included data from 169 respondents, of whom 62 were men and 107 were women. The mean age of the entire sample was 55.21±13.47; the mean duration of the disease was 11.79±8.16. All participants signed an informed consent prior to the study as approved by the Ethics Board of Moscow State University of Medicine and Dentistry. The dataset was collected based on the V.A. Nasonova Scientific Research Institute of Rheumatology and the A.S. Loginov Moscow Clinical Research Center. The study was approved by both institutions.

**Tab. 1. The mean age and mean duration by group with significant differences are as follows**

|  | Group 1<br>N=58<br>m ± SD | Group 2<br>N=45<br>m ± SD | Group 3<br>N=66<br>m ± SD | Significance of differences |
|---|---|---|---|---|
| Age | 56,93 ± 13,63 | 50,87 ± 13,53 | 57,32 ± 12,70 | - |
| Duration of illness | 11,45 ± 8,12 | 10,02 ± 7,84 | 13,17 ± 8,25 | p1-p3=0,025 |

**Tests:**

1. Brief disease perception questionnaire [28];
2. Questionnaire of coping with difficult life situations [26];
3. Hospital Anxiety and Depression Scale [29];
4. Five-factor mindfulness questionnaire [11];

5. Treatment adherence questionnaire [18].

**Statistical analysis:**

Statistical data processing was performed using the SPSS Statistics program (Vers. 23). Analysis of variance was used to compare mean values between groups. Differences were considered statistically significant at p<0.05. Factor analysis was performed using the principal component method, as well as the Varimax factor rotation method with the Kaiser-Meyer-Olkin test. Data clustering was performed using the k-means method (with ANOVA analysis conducted).

**Funding:**

# Results

The first step in analyzing the results was to identify basic profiles and types of perception of type 2 diabetes among patients. When identifying basic profiles, it was decided to rely not on the traditional nosological criterion, but on a psychological criterion. The factorization was based on the patient's perception of the threat of the disease and their beliefs about its controllability, which were measured using the "Brief Illness Perception Questionnaire". The available data for each patient were clustered according to the degree of expression of the identified factors.

Two statistically significant factors were found to highlight illness perception profiles:

**Tab. 2. Results of factor analysis.**

|  | Component | |
|---|---|---|
|  | 1 | 2 |
| Perception of disease threat | **0,848** | 0,274 |
| Concern | **0,779** | 0,036 |
| Identity | **0,750** | -0,035 |
| Consequences | **0,744** | -0,025 |
| Emotional representation | **0,731** | 0,059 |
| Timeline | **0,462** | -0,129 |
| Coherence | 0,063 | **0,797** |
| Personal control | -0,003 | **0,775** |
| Treatment control | -0,034 | **0,774** |

Profile 1 (Factor 1) is named "Perception of Threat of Illness" and has the following components of illness perception:

1. Perception of disease threat - something perceived by the patient as a physical, emotional or social risk that they believe has the potential to affect, or is already affecting, their somatic and mental health;

2. Concern - the patient's level of preoccupation with the disease;

3. Identity - the ability to recognize painful sensations and classify them as symptoms of a given disease;

4. Consequences - the patient's perceptions of the effects of the disease (biological, psychological or social) that will affect their quality of life;

5. Emotional representation - the level of distress experienced by the patient;

6. Timeline - patient's perceptions of the duration of the disease.

Profile 2 (Factor 2), Perceptions of controllability of illness and treatment, has three components:

1. Coherence - the patient's subjective assessment of his or her own understanding of the etiology and pathogenesis of the disease.

2. Personal control - the patient's perceptions of the possibility of cure or at least successful control of the disease

3. Treatment control - the patient's perceptions of the effectiveness of therapy, the work of medical personnel, and his or her own efforts to treat the disease.

According to the results of clustering we identified three groups of patients, each of which corresponds to one of three types of disease perception: type 1 (cluster 1) "Formed threat of disease and negative perceptions of disease control and treatment" - group 1; type 2 (cluster 2) "Unformed threat of disease and neutral perceptions of disease control and treatment" - group 2, type 3 (cluster 3) "Formed threat of disease and positive perceptions of disease control and treatment" - group 3. The graphical representation of the types of disease perception is presented below.

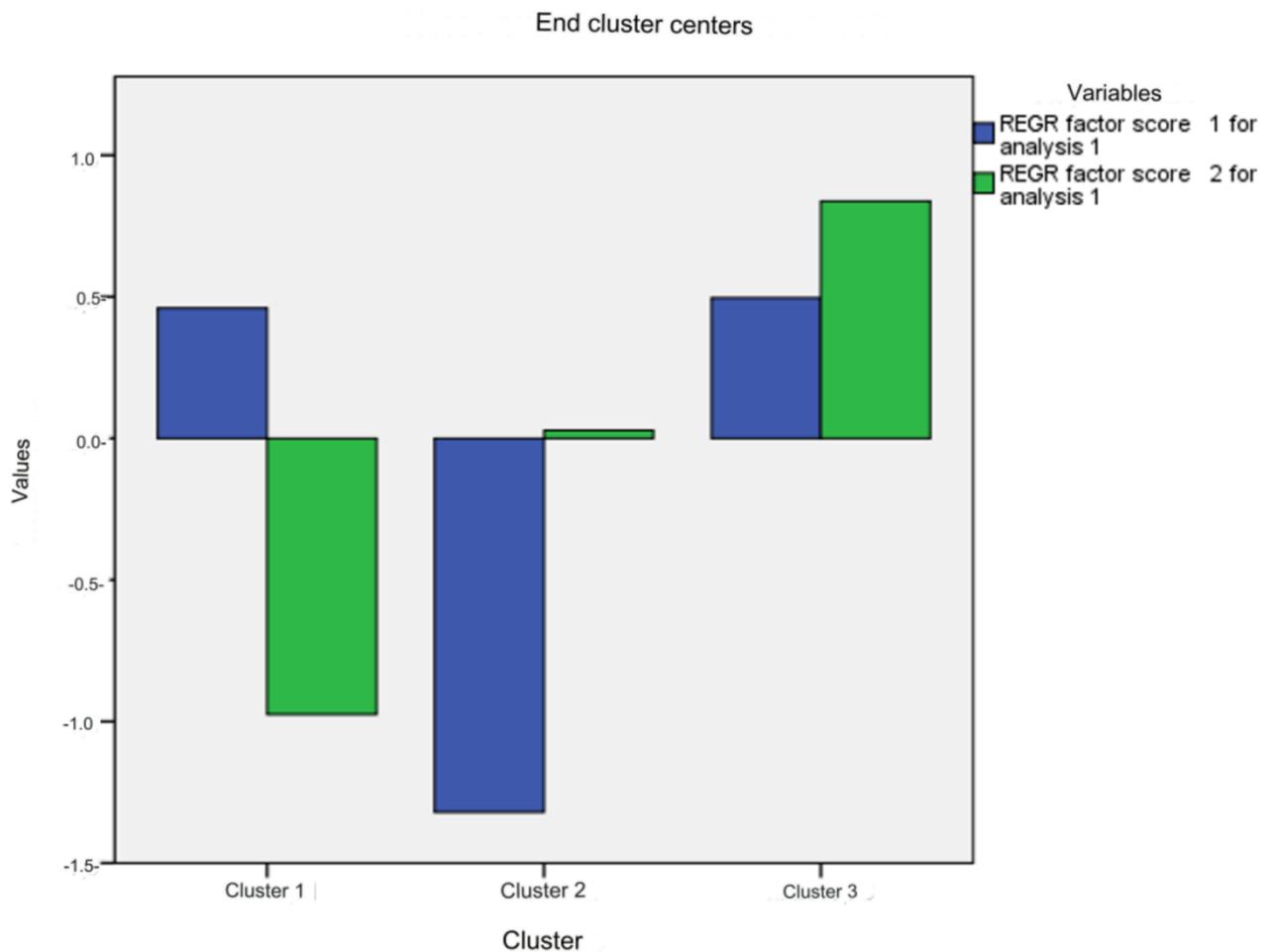

Fig. 1. The graphical representation of the types of disease perception.

Type 1 "Formed disease threat and negative beliefs about disease control and treatment" - patients who perceive type 2 diabetes as a marked threat, but are convinced that the disease is difficult to control and the treatment provided is not effective enough.

Type 2 "Unformed threat of disease and neutral ideas about disease control and treatment" - the disease is not perceived by the patient as a threat to life, ideas about disease control and treatment are nominal. In this case, the patient can follow some recommendations of doctors, but with great skepticism about the effectiveness of therapy, or generally doubting its relevance.

Type 3 "Formed threat of disease and positive perceptions of disease control and treatment" - these patients perceive type 2 diabetes as a pronounced threat. At the same time, they make efforts to control the disease, following doctors' prescriptions and evaluating treatment as a necessary measure to cope with the disease.

# Comparative analysis of the described types of perception of type 2 diabetes.

**Tab. 3. Results of a comparative study of coping behavior strategies.**

|  | Group 1<br>N=58<br>m ± SD | Group 2<br>N=45<br>m ± SD | Group 3<br>N=66<br>m ± SD | Significance of differences |
|---|---|---|---|---|
| Confrontation | 48,24 ± 11,01 | 52,84 ± 10,46 | 55,25 ± 12,31 | p1-p3 = 0,005 |
| Distancing | 50,84 ± 11,97 | 53,71 ± 10,82 | 52,76 ± 10,08 | - |
| Self-control | 47,27 ± 13,85 | 47,53 ± 10,46 | 50,22 ± 9,64 | - |
| Searching for social support | 47,71 ± 10,65 | 48,33 ± 8,39 | 52,13 ± 9,09 | p1-p3 = 0,013<br>p2-p3 = 0,029 |
| Acceptance of responsibility | 47,47 ± 10,78 | 47,07 ± 8,35 | 49,56 ± 7,90 | - |
| Escape-avoidance | 51,87 ± 10,15 | 53,91 ± 9,82 | 55,11 ± 9,25 | - |
| Planning for problem solving | 46,42 ± 12,11 | 46,91 ± 11,51 | 49,27 ± 7,76 | - |
| Positive revaluation | 47,22 ± 12,50 | 48,56 ± 9,87 | 50,62 ± 10,78 | - |

In this methodology, all values are within the norm, indicating a moderate and balanced use of coping behavior strategies in each group. The table shows that the average values of group 3 are overwhelmingly higher than in groups 1 and 2. Thus, it can be said that there is a tendency towards more active coping behavior in group 3. Significant differences were found in the scales "Confrontation" (p1-p3 = 0.005) and "Search for social support" (p1-p3 = 0.01; p2-p3 = 0.029), which suggests that group 3 is more prone to strategies of resistance and search for help from outside [11,17,18, 25].

**Tab. 4. Results of comparative analysis of anxiety and depression.**

|  | Group 1<br>N=58<br>m ± SD | Group 2<br>N=45<br>m ± SD | Group 3<br>N=66<br>m ± SD | Significance of differences |
|---|---|---|---|---|
| Anxiety | 7,35 ± 3,35 | 7,86 ± 4,27 | 7,68 ± 4,34 | - |
| Depression | 6,82 ± 3,37 | 8,19 ± 3,73 | 7,37 ± 4,39 | p1-p2 = 0,050 |

According to the methodology, all values are within the normative corridor, but are located close to the upper limit. The exception is the scale "Depression" in group 2 - here the value has an indicator of subclinical severity. In groups 1 and 3 the anxious radical is more pronounced, which coincides with the result of clustering, where the level of disease threat is higher in groups 1 and 3. The depressive radical is more consistent for group 2. A significant difference was found for the Depression scale (p1-p2 = 0.050). The results show that group 2 has the highest intensity of experience, with a greater expression of the depressive component, but according to the clustering this group has the lowest scores on the level of threat of illness. Because of this, we have no reason to assume that the results in this group directly depend on the experience of disease threat.

**Tab. 5. Results of comparative analysis of treatment adherence.**

|  | Group 1 N=58 m ± SD | Group 2 N=45 m ± SD | Group 3 N=66 m ± SD | Significance of differences |
|---|---|---|---|---|
| *Treatment adherence* | 5,76 ± 1,58 | 4,73 ± 1,89 | 4,91 ± 2,17 | p1-p2 = 0,050 p1-p3 = 0,036 |

This technique showed that the level of adherence to treatment in all groups is below normal, indicating nonadherence of patients in all 3 comparison groups. The scores of groups 2 and 3 are statistically significantly lower than group 1 (p1-p2 = 0.050; p1-p3 = 0.036), while the factor "Perception of controllability of illness and treatment" of groups 2 and 3, on the contrary, is higher than group 1. This indicates the belief of groups 2 and 3 that their illness is controlled by some other means than the therapy prescribed by the doctor. Overall, the results indicate a general distrust of therapy and its underestimation by patients [27].

**Tab. 6. Results of comparative analysis of the level of mindfulness in the surveyed groups.**

|  | Group 1 N=58 m ± SD | Group 2 N=45 m ± SD | Group 3 N=66 m ± SD | Significance of differences |
|---|---|---|---|---|
| Description | 29,19 ± 6,24 | 25,66 ± 4,91 | 23,58 ± 5,76 | p1-p2 = 0,005 p1-p3 = 0,002 |
| Conscious activity | 27,16 ± 6,25 | 24,46 ± 5,89 | 24,53 ± 4,90 | p1-p2 = 0,050 |
| Non-judgmental attitude to experience | 24,86 ± 5,99 | 26,37 ± 6,17 | 27,84 ± 4,75 | p1-p3 = 0,005 |
| Unresponsiveness | 20,43 ± 4,09 | 21,37 ± 4,41 | 18,79 ± 3,87 | p2-p3 = 0,040 |

According to the methodology, on the scale "Description" groups 2 and 3 have indicators below the normative ones. For other scales the results correspond to the norm. Table 5 shows that group 1 has higher scores on the "Description" and "Conscious activity" scales; group 2 - on the

"Non-response" scale; group 3 - on the "Non-evaluative attitude to experience" scale. Many significant differences were obtained for all scales of the methodology, but group 1 more often has statistically significant differences with respect to groups 2 and 3, which indicates a certain isolation of the first group from the others in this particular methodology [7,9,28].

## Discussion

This study is a continuation of our paradigm formulated in a study of illness perception and adaptation to illness in patients with immunoinflammatory rheumatic diseases [2], and also continues the general direction of our research.

If we describe the results of the study conceptually, we can notice that the results of some of them "overlap" with each other. Namely, the results of the methods "Brief Illness Perception Questionnaire" [28] (it was used for case factorization and clustering), "Coping with Difficult Life Situations Questionnaire" [26] and "Five-Factor Awareness Questionnaire" [11] have some common configuration of results. Overlaying the results of the methods on clustering, we see the following:

Group 1, having a high factor of threat of illness, but a reduced value in the perceptions of control, shows more moderate use of all coping strategies and at the same time more pronounced skills of active awareness of the situation. Conditionally, this group can be called "experiencing observer" - there is tension regarding the fact of illness, but for one reason or another there is no active struggle with it, and coping is aimed at realizing the situation.

Group 3 is opposite to group 1. It has a high index on the factor of threat of illness, but at the same time, it also has a high value on the factor of perceptions of control. There is also more active coping with the illness in combination with a more pronounced mindfulness skill "Non-judgmental attitude towards experience". Patients feel threatened by illness, struggle with it and accept the experience of coping, both positive and negative. This group is conventionally termed the "experiencer".

If group 1 and group 3 are "two sides of the same coin", group 2 is its "edge". In contrast to the above-mentioned groups, group 2 does not feel a pronounced threat of illness, and perceptions of disease control and treatment can be interpreted as neutral. The indicators on coping strategies have intermediate values between group 1 and 3, and among the skills of mindfulness the skill of "non-response" stands out. This group we conditionally called "wary performer". This category of patients accepts the fact of the disease and can even follow some recommendations of doctors, but while the interaction with the situation of the disease is mechanical, the patient as if has not yet reflected the topic of the disease and therefore the processes of its perception have not yet been launched.

It is important to note that the HADS [29] and MMAS-8 [18] techniques stand alone in this theoretical model. Factorization showed different levels of threat perception among patients, but anxiety and depression levels were comparable to each other and within normal limits. Such a result probably indicates the presence of extraneous variables affecting anxiety and depression scores in the comparison groups. Regarding the MMAS-8, the situation is similar: despite different measures of perceptions of disease control and treatment, patients have similar values of (low) treatment adherence.

It is also important to emphasize that the similarities and differences of the groups described above are schematic and pretend to describe only some tendency. In particular, the descriptions of coping strategies are conditional, since all comparison groups demonstrated moderate and balanced preferences for certain coping strategies.

However, the results of this study provide an opportunity to identify the main targets for psychological work. For the first type, this target is likely to be building trust in doctors and their methods of work, working with cognitive errors associated with misconceptions about type 2 diabetes therapy. For the second type of patient, the focus should be on developing an adequate critical attitude towards the disease. For the third type, psychological work should be focused on prevention or reduction of burnout, support and containerization of negative experiences, as well as help in finding psychological resources for coping with the disease.

In general, a peculiarity of type 2 diabetes is that the patient does not immediately notice the manifestations of this serious disease. For some, the final realization and recognition of the disease comes in the diabetic foot department after the onset of necrosis of tissues of the lower extremities or retinopathy. Therefore, the timely formation of stable commitment to treatment of these patients is one of the primary tasks of the clinical psychologist, on which depends the success of all further therapy.

## Conclusion

1. Type 2 diabetes perception can be considered in the context of disease threat perception and disease control and management. Accordingly, the first profile "Perception of disease threat" includes the following characteristics: "Threat of disease", "Consequences of disease", "Concern about disease", "Emotional response to disease", "Identification of disease manifestations", "Perception of disease course". This profile reflects how dangerous the patient sees his illness. The second profile "Perception of controllability of the disease and treatment" includes such components as "Controllability of the disease", "Controllability of treatment" and "Understanding of the disease".

It characterizes the patient's subjective assessment of the possibility of controlling the disease and the effectiveness of its treatment.

2. Depending on the degree of expression of the described profiles, three types of psychological perception of type 2 diabetes were identified: "Unformed control of the disease" (type 1), "Unformed threat of the disease" (type 2), "Formed threat and control of the disease" (type 3).

3. Reliance on the psychological criterion of analysis made it possible to identify profiles and types of disease perception, as well as to identify the main targets of psychological work corresponding to this or that type.